LA-UR-

*Title:*

*Author(s):*

*Intended for:*

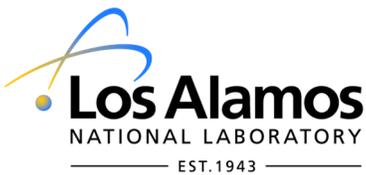

# Proton-induced cross sections relevant to production of $^{225}$Ac and $^{223}$Ra in natural thorium targets below 200 MeV

J.W. Weidner*, S.G. Mashnik, K.D. John, F. Hemez,
B. Ballard, H. Bach, E.R. Birnbaum, L.J. Bitteker, A. Couture, D. Dry, M.E. Fassbender, M.S. Gulley,
K.R. Jackman, J.L. Ullmann, L.E. Wolfsberg,
and F.M. Nortier[1]

*Los Alamos National Laboratory, Los Alamos, New Mexico, 87545, USA*

**Abstract**

Cross sections for $^{223,225}$Ra, $^{225}$Ac and $^{227}$Th production by the proton bombardment of natural thorium targets were measured at proton energies below 200 MeV. Our measurements are in good agreement with previously published data and offer a complete excitation function for $^{223,225}$Ra in the energy range above 90 MeV. Comparison of theoretical predictions with the experimental data shows reasonable-to-good agreement. Results indicate that accelerator-based production of $^{225}$Ac and $^{223}$Ra below 200 MeV is a viable production method.

* Corresponding author. Tel.: +1 505-665-6956.
  *E-mail addresses*: john.weidner@us.army.mil (J.W. Weidner), meiring@lanl.gov (F.M. Nortier).
  [1] Principle investigator.

## 1. Introduction

Interest in the use of alpha-emitting isotopes for therapeutic treatment has grown steadily over the past decade (Apostolidis et al., 2005, Couturier et al., 2005). Two such isotopes with strong therapeutic potential are $^{225}$Ac and $^{223}$Ra. These and other alpha-emitting candidates tend to be in short supply, however, due to the unique manner in which they are produced or generated. For example, the annual worldwide production of $^{225}$Ac is on the order of 1 Ci, which is far below the anticipated demand should full-blown clinical trials and R&D be initiated (Norenberg et al., 2008). Hence, the investigation of accelerator production routes for $^{225}$Ac has been encouraged by the National Science Advisory Committee's Isotopes subcommittee, which was appointed by the U.S. Department of Energy (DOE) Office of Science, Office of Nuclear Physics (NSAC-I, 2009). Additionally, recent IAEA reports (Nichols et al., 2011, Capote Noy and Nortier, 2011) emphasize the need for additional cross section measurements of $^{225}$Ac, its parent $^{225}$Ra, and $^{223}$Ra. This work investigates accelerator-based production of $^{225}$Ac and $^{223}$Ra by proton irradiation of thorium targets in the energy range below 200 MeV. Cumulative cross sections for these and other isotopes related to their production were measured and are reported.

The work reported here represents the second phase of a two-phase investigation into the high-energy accelerator production of $^{225}$Ac and $^{223}$Ra. The first phase measured the production cross sections for production via spallation at 800 MeV, and the results have recently been published (Weidner et al., 2011). The reader is referred to this article for a more detailed description of current production methods for these isotopes and their clinical uses. This article focuses on the results of the cross section measurements applicable to high-energy production facilities employing beam energies of up to 200 MeV.

## 2. Experimental technique

### 2.1 Irradiation

All elements of this work occurred at Los Alamos National Laboratory. A two step approach was used to measure the desired cross sections below 200 MeV. In the first step, a target consisting of six foils separated by aluminum degraders was irradiated with protons having an initial energy of 197.1 ± 0.3 MeV (referred to as a 200 MeV beam) in the Target 2 area (Blue Room) of the Weapons Neutron Research facility. The dual-energy capability of the 800 MeV side-coupled cavity linear accelerator of the Los Alamos Neutron Science Center (LANSCE) facility was re-established to conduct these lower energy experiments in parallel with our 800 MeV experiments. The 200 MeV irradiation intended to measure cross sections down to proton energies of approximately 100 MeV. In the second step, a similar stacked foil target was irradiated at the LANSCE Isotope Production Facility, which receives its 100.0 ± 0.3 MeV proton beam from a drift tube linear accelerator located upstream of the side-coupled cavity accelerator (Lisowski and Schoenberg, 2006).

In each irradiation, a target stack was comprised of six foil packets, each containing two thorium foils and one aluminum monitor foil. Natural thorium foils of 99.7% purity were obtained from Goodfellow Corporation (Oakdale, PA). The foils measured approximately 2.5 cm x 2.5 cm and had thicknesses ranging from 60.5 to 70.5 mg/cm$^2$. A similar sized aluminum beam current monitoring foil of 99.9% purity with a thickness of approximately 65 mg/cm$^2$ was incorporated downstream of each set of thorium foils. All foils were sandwiched between single layers of Kapton tape with a thickness of 25 μm, which served as a catcher foil for recoil ions. The combination of the two irradiation experiments yielded cross section measurements at the following 11 energies: 194.5 ± 0.3, 178.3 ± 0.7, 160.7 ± 1.0, 141.8 ± 1.3, 120.9 ± 1.6, 97.0 ± 2.0, 90.8 ± 0.4, 81.7 ± 0.6, 72.8 ± 0.7, 64.9 ± 0.9, and 56.3 ± 1.1 MeV. Unfortunately, the non-destructive gamma spectroscopy techniques utilized were not sensitive enough to measure the cross section for any isotope of interest at 46.4 MeV. Both foil stacks were irradiated for approximately one hour without interruption; the average beam current was 71.2 nA for the 200 MeV irradiation and 126.8 nA for the 100 MeV irradiation.

2.2  Gamma spectroscopy

This work utilized nondestructive gamma spectroscopy of the foils to quantify the activity of each isotope of interest. The foils were prepared for gamma spectroscopy several hours after end of bombardment (EOB). The thorium foils were counted on an ORTEC GEM10P4-70 detector with a relative efficiency of 10%, while the aluminum foils were counted on a Princeton Gamma-tech lithium-drifted germanium Ge(Li) detector with a relative efficiency of 13.7%. Both detectors were well shielded and calibrated using NIST-traceable gamma calibration sources. The thorium foils were counted more than 35 times over a period of several months and the decay curves of all isotopes of interest were closely followed to ensure proper identification and to evaluate any possible interferences. The aluminum foils were counted approximately 12 times within the first week after EOB to monitor the $^{24}$Na decay curve, followed by a minimum of three 8-hour counts several weeks later to quantify the $^{22}$Na activity at EOB.

2.3  Data analysis

Analysis of the gamma spectra was conducted utilizing the same LANL-specific analysis codes as in our 800 MeV experiment; gamma ray energies and intensities as listed on the Lund/LBNL Nuclear Data Search website (Chu et al., 1999) were used to determine the activity of the relevant isotopes and are listed in our 800 MeV report (Weidner et al., 2012). The incident flux on each stack was determined from the activity produced in the Al monitor foils using the well-characterized $^{27}$Al(p,x)$^{22}$Na reaction cross section. All monitor foil data from a particular stack were considered simultaneously in order to obtain the best overall agreement with the relevant part of the measured monitor excitation function. Cross section values as measured by Steyn et al. (1990) were used for the energy range between 200 and 100 MeV, while the IAEA recommended values (Tarkanyi et al., 2001), retrieved from the IAEA charged particle cross section database, were used for measurements below 100 MeV. Uncertainties were attributed to the IAEA values by interpolating the uncertainties from the measurements of Steyn et al. below 100 MeV. The $^{27}$Al(p,α)$^{24}$Na reaction was also considered as a monitor reaction. However, the proton fluence calculated from this reaction was consistently lower than that obtained from the $^{27}$Al(p,x)$^{22}$Na reaction, with a 40% discrepancy noted for the lowest energy foil in the 100 MeV target stack. Since the discrepancy is believed to result from secondary neutron contributions via other competing reactions, e.g., $^{27}$Al(n,α)$^{24}$Na, this monitor reaction was rejected. An MCNP6 (Goorley et al., 2011) simulation was created for both the 200 and 100 MeV target stacks and used to calculate the proton energy distribution at each foil. To account for proton straggling, these models incorporated a tally of the proton energy distribution in 0.5 MeV increments. The uncertainties in the very Gaussian-like proton energy distributions are taken as one standard deviation.

Numerous interferences with the gamma signatures of $^{225}$Ac, $^{223}$Ra, $^{225}$Ra, and $^{227}$Th were observed. Activities for $^{225}$Ac and $^{225}$Ra were quantified based upon the measured activity of $^{221}$Fr, while $^{223}$Ra and $^{227}$Th were quantified based upon the activity of $^{211}$Bi. Thorium-227 was also quantified directly by measuring its 235.971 keV gamma. The $^{227}$Th activities at EOB obtained from these two methods agreed to within 5%.

The activities of all isotopes at EOB, and their 1 σ uncertainties, were calculated by performing a one million sample Markov chain Monte Carlo (MCMC) analysis of each measured decay curve. Fist, the data from the measured $^{221}$Fr, $^{211}$Bi or $^{227}$Th decay curve is loaded into the code, along with an initial guess of the parameters to be fit (in this case, the EOB activity of the parent radioisotopes). The initial guesses were obtained from a least squares fit of the measured data. The MCMC code then takes a random walk away from the initial guess (or guesses when fitting two activities simultaneously, such as $^{225}$Ac and $^{225}$Ra) and selects a new value for each parameter to be fit. This random walk occurs within a boundary of values predefined by the user. When new sample values are obtained, a theoretical decay curve is created for the isotope that was measured ($^{221}$Fr, $^{211}$Bi, or $^{227}$Th). This theoretical curve is then assessed against how well it passes through the error bars of the measured data points. If the sampled values lead to a model that accurately mimics the measured data (as determined by a preset variance within the code), those values are saved and the random walk continues from those last sampled parameter values. If the model is rejected, the random walk continues from the last accepted parameter values. Once the desired number of accepted models has been achieved, the code terminates the sampling routine and calculates the mean and standard deviation of the accepted sample distribution.

The total uncertainty in each measured data point used in the MCMC analysis was the quadrature sum of the estimated uncertainties in detector calibration (2.1%), counting geometry (1%), and the gamma peak area (2-10%). The uncertainties in the $^{225}$Ac and $^{227}$Th EOB activities inferred by the MCMC analysis were below 2% for all energy points, resulting in cross section values with relatively small error bars. Conversely, the uncertainties in the $^{223,225}$Ra activities at EOB inferred by the MCMC analysis approached 60% at low energies, thereby leading to larger uncertainties in the cross section values below 100 MeV. Longer irradiation times, increased beam current or chemical separation of the isotopes of interest from the thorium matrix could improve the accuracy of these cross section measurements.

The total uncertainties in the final cross section values were calculated by summing individual contributing uncertainties from the foil thickness (<1%), integrated proton current (6.4-7.3%), and radioisotope activities at EOB in quadrature. The uncertainty associated with each cross section value is expressed as a 1 $\sigma$ confidence level.

## 3.     Results and Discussion

### 3.1     Cumulative Cross Sections

The measured cumulative cross sections for the production of $^{223,225}$Ra, $^{225}$Ac, and $^{227}$Th by proton irradiation of thin, natural thorium foils with protons below 200 MeV are shown in Table 1. Figures 1-4 graphically illustrate the comparison of both the theoretical values and previously measured values to the results of this work.

### 3.1.1     $^{232}$Th(p,x)$^{225}$Ac and $^{232}$Th(p,x)$^{225}$Ra reactions

Our measured and theoretical $^{232}$Th(p,x)$^{225}$Ac cross sections account for the decay of $^{229}$Pa and $^{225}$Th adding to the measured activity of $^{225}$Ac. The three measurements by Zhuikov et al. (2011) are in excellent agreement with our results, as are the data by Ermolaev et al. (2012) – particularly those measurements above 125 MeV (using a proton beam of initial energy 158.5 MeV) and below 75 MeV (using a proton beam of initial energy 100.1 MeV). Below 100 MeV, the Ermolaev et al. measurements display a slight, self-reported discrepancy in the two data sets. The single data points measured by Lefort et al. (1961) and Titarenko, el al. (2002 and 2003) are also in very good agreement with our data. Conversely, the measurements by Gauvin et al. (1962) and Gauvin (1963) show discrepancies. With the exception of data at very low energies, their measurements are lower by approximately a factor of two.

In the case of the $^{232}$Th(p,x)$^{225}$Ra production route, measured cross sections for the formation of $^{225}$Ra include contributions from the decay of $^{225}$Fr. The measurements by Hogan et al. (1979) are in very good to excellent agreement with our data, especially below 75 MeV, while the data reported by Lefort et al. (1961), appear to be larger by more than a factor of two.

### 3.1.2     $^{232}$Th(p,x)$^{223}$Ra and $^{232}$Th(p,x)$^{227}$Th reactions

Both the theoretical and measured cross sections for the $^{232}$Th(p,x)$^{223}$Ra reaction include contributions from the decay of $^{223}$Fr and $^{223}$Ac. While the single measurement by Lefort is only slightly lower than our data, the values obtained by Hogan et al. are consistently lower by approximately a factor of two.

The $^{232}$Th(p,x)$^{227}$Th cross sections include contributions from the decay of $^{227}$Pa. In the energy range above 110 MeV, the measurements by Zuikov et al., Titarenko et al., and Lefort et al., are in excellent agreement with our values. Between 60-100 MeV, all other measurements (except for the 100 MeV measurement by Titarenko et al.) are consistently lower than the values measured in this work and also do not reflect the broad resonance peak of nearly 50 mb indicated by the data reported here.

### 3.1.3 $^{232}$Th(p,x)$^{227}$Ac reactions

Cross sections for the long-lived beta emitter $^{227}$Ac were not measured. This required chemical separation of actinium from the thorium matrix, which was beyond the scope of the present work. In order to estimate the contribution of the $^{227}$Ac impurity to the $^{225}$Ac produced in a thorium target, the cross section measurements obtained from gamma spectroscopy by Ermolaev et al. (2012) were used for proton energies below 141 MeV. Theoretical estimates using ALICE2010 (Blann et al., 2010) are in very good agreement with the Ermolaev et al. data, and were used in the absence of measurements above 141 MeV (Fig. 5).

### 3.2 Theoretical predictions

Theoretical independent formation cross section values were calculated for isotopes relevant to the production of $^{223,225}$Ra, $^{225}$Ac and $^{227}$Th in thin thorium targets. These calculations utilized ALICE2010 (Blann et al., 2010), as well as the MCNP6 code with three different event-generators: the Cascade-Exciton Model as implemented in the code CEM03.03 (Mashnik et al., 2008), the Bertini+MPM+Dresner+RAL event-generator (Bertini, 1969; Prael and Bozoian, 1988; Dresner, 1981; Atchison 1980), and the Liege intra-nuclear cascade model INCL (Boudard et al., 2002) merged with the ABLA evaporation/fission model (Junghans et al., 1998). Standard default values of all parameters were used in each model. Cumulative theoretical cross sections for the isotopes of interest were derived from the independent cross sections by taking into account the appropriate contributions from the parent isotopes as identified above.

The theoretical models generally show agreement to within a factor of two of the experimental values measured in this work; however, no single model provides a set of theoretical values that consistently matches all of the cross sections reported here. This trend is most apparent when the models are compared to our $^{225}$Ac cross section measurements. The ALICE2010 predictions are a factor of 2-3 less than our measurements for all energies below 200 MeV, while the remaining models overpredict our measurements above 100 MeV. For $^{223}$Ra, the CEM model best predicts our cross sections below 100 MeV, but is the least accurate predictor above 100 MeV. The ALICE2010 model closely follows the slope of the excitation function as it increases with energy, though its predicted values are generally between 0.5-1.0 mb less than our measurements. None of the four theoretical models agree with measurements for the production of $^{225}$Ra across any significant energy range. Above 125 MeV, all models overpredict our measurement by factors of 2-4, with the CEM and ALICE2010 models providing the most accurate estimates in that energy range. Lastly, the CEM model reproduces very well the overall shape of our measured $^{227}$Th excitation function, though both it and the Bertini model forecast a steeper rise in the cross section value occurring just beyond the reaction threshold. Above 110 MeV, the INCL and CEM models are in very good agreement with our $^{227}$Th data. This fluctuation in theoretical predictions demonstrates the need for further refinement of the nuclear models and that experimental cross section measurements are still required to accurately predict isotope production yields.

### 3.3 Production rates and yields

For comparison purposes, Table 2 lists the instantaneous production rate and total EOB yield of $^{223,225}$Ra, $^{225,227}$Ac and $^{227}$Th for one of several practical irradiation scenarios at the LANL Isotope Production Facility (IPF) and the Brookhaven Linac Isotope Producer (BLIP). Assumptions include an uninterrupted 10-day irradiation, a thorium production target with a thickness of 5 g/cm$^2$, a 100 MeV, 250 µA proton beam at IPF and a 200 MeV, 100 µA proton beam at BLIP. All isotopic yields from such an irradiation, but particularly those for $^{225}$Ac and $^{227}$Th, would provide a significant addition to the current annual worldwide yield of $^{225}$Ac and $^{223}$Ra. For instance, coordinated production runs at IPF and BLIP could increase the present annual worldwide supply of $^{225}$Ac by up to sixty-fold, even if regularly scheduled annual downtime of each accelerator is accounted for. Additional thorium targets can also be utilized downstream of the initial target to further increase the yield at each facility.

#### 3.3.1 Production of $^{225}$Ac

In addition to production of $^{225}$Ac directly via nuclear reactions and the decay of short-lived parents, or indirectly by the decay of its longer-lived parent $^{225}$Ra as reported here, a third, lower energy production route is also possible via the path $^{229}$Pa (T$_{½}$ = 1.50 d) → $^{229}$Th (T$_{½}$ = 7,340 y) → $^{225}$Ra. Cross sections for the latter path are not reported since the measurements in this work could quantify neither $^{229}$Pa, due to gamma interferences, nor $^{229}$Th, due to its very low activity.

Depending upon the facility, the illustrated 10-day irradiation anticipates a directly produced $^{225}$Ac yield of up to 2.0 Ci, doubling the current annual worldwide $^{225}$Ac supply in a single production run. Though other undesirable isotopes of actinium would also be produced, $^{227}$Ac is the only one with a half-life greater than that of $^{225}$Ac. The co-production of this long-lived impurity is a concern. However, the predicted $^{227}$Ac activity typically represents 0.2% or less of the total activity of the $^{225,227}$Ac combination. Additional research into the biological effect of $^{227}$Ac within a $^{225}$Ac carrier is needed to ultimately determine if the presence of this impurity would preclude accelerator-produced $^{225}$Ac as a viable option for targeted alpha therapy. On the other hand, the presence of this impurity is not likely to diminish the value of produced $^{225}$Ac for use in a $^{213}$Bi generator.

Though more than an order of magnitude less than the $^{225}$Ac yield, the $^{225}$Ra yield from the production example in Table 2 is still significant and could be utilized as a pure $^{225}$Ac generator. Although $^{227,228}$Ra are also created by this production method, they will act as very small sources for radio-actinium impurities through decay. The $^{227}$Ra has a 42.2 min half-life, but decays to the long-lived impurity $^{227}$Ac. Conversely, the relatively long-lived $^{228}$Ra has a half-life of 5.75 years, but decays to the short-lived isotope $^{228}$Ac (T$_{½}$ = 6.15 h), which in turn beta-decays to the alpha emitter $^{228}$Th (T$_{½}$ = 1.9 y). The impact of $^{227}$Ra production can be mitigated by delaying the chemical separation of radium until after the majority of this isotope has decayed into $^{227}$Ac. Given its 5.75 y half-life, the dose contributed by $^{228}$Ra is expected to be very low.

#### 3.3.2 Production of $^{223}$Ra

Two routes for $^{223}$Ra production are evident from our results. First, the direct production of $^{223}$Ra via nuclear reactions and decay of short-lived parent nuclides is expected to yield several hundred milliCuries as shown in Table 2. Of the contaminant isotopes of radium that are co-produced, only three are of concern: $^{225,226,228}$Ra. The 1600 y half-life of $^{226}$Ra may make its presence tolerable; further research into the biological fate of radium is needed to determine if the presence of $^{225,228}$Ra is acceptable. Despite the presence of these impurities, accelerator-based production of $^{223}$Ra as a generator of $^{211}$Pb appears viable since only low levels of daughter products of the contaminants are assumed to be discharged from the generator.

Second and more significantly, the example irradiation is expected to create nearly 9 Ci of $^{227}$Th at the LANL facility (IPF) and nearly 2 Ci at Brookhaven (BLIP). Although other alpha-emitting isotopes of thorium would also be produced, their half-lives are all either much shorter or much longer than that of $^{227}$Th. Therefore, careful timing of the target's chemical processing should lead to the recovery of a relatively high quality $^{227}$Th product that could serve as a multi-Curie generator for pure $^{223}$Ra.

## 4. Summary and Conclusion

The cross sections measured in this work provide a significant addition to the published $^{223,225}$Ra data above 90 MeV and improve the database of $^{225}$Ac and $^{227}$Th values, particularly above 150 MeV. Comparison of theoretical model predictions to our measured values shows generally good agreement, with all of our measurements positioned near the median of the predicted values. No one model was shown to be consistently more accurate than another.

Yield estimates indicate that multi-Curie quantities of $^{225}$Ac and $^{223}$Ra (the latter as obtained from a $^{227}$Th generator) can be co-produced in a single 10-day irradiation using the intense proton beams available at IPF and BLIP. Several hundred mCi of pure $^{225}$Ac can also be obtained from the $^{225}$Ra produced. Proper timing of the target's chemical processing should provide a multi-Curie $^{227}$Th product suitable for use in a generator for the production of pure $^{223}$Ra. Though accelerator-based production of $^{225}$Ac also leads to the production of $^{227}$Ac, the activity of this impurity is expected to be less than 0.2% of the overall activity of the $^{225,227}$Ac combination and may prove tolerable for therapy applications with further research. Additionally, the presence of impurities should have negligible impact on the use of accelerator-produced $^{225}$Ac and $^{227}$Th as $^{213}$Bi and $^{211}$Pb generators, respectively.

**Acknowledgements**

We gratefully acknowledge funding by the United States Department of Energy Office of Science via an award from The Isotope Development and Production for Research and Applications subprogram in the Office of Nuclear Physics. We are also very thankful for the technical assistance provided by members of the LANL C-NR and C-IIAC groups, members of the LANSCE AOT-OPS group, and the LANSCE-WNR staff.

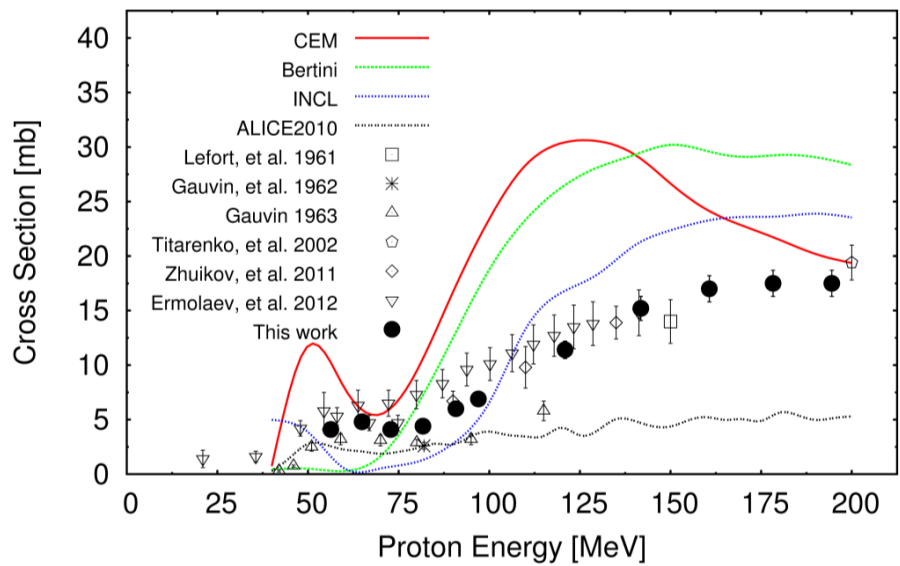

**FIG. 1.** Experimental and theoretical cumulative cross sections for the formation of $^{225}$Ac by the proton bombardment of thorium.

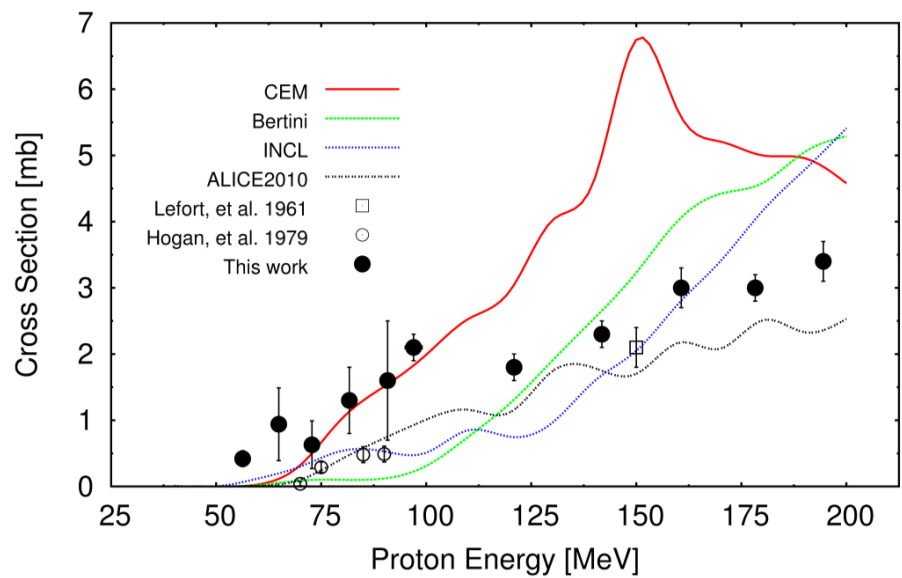

**FIG. 2.** Experimental and theoretical cumulative cross sections for the formation of $^{223}$Ra by the proton bombardment of thorium.

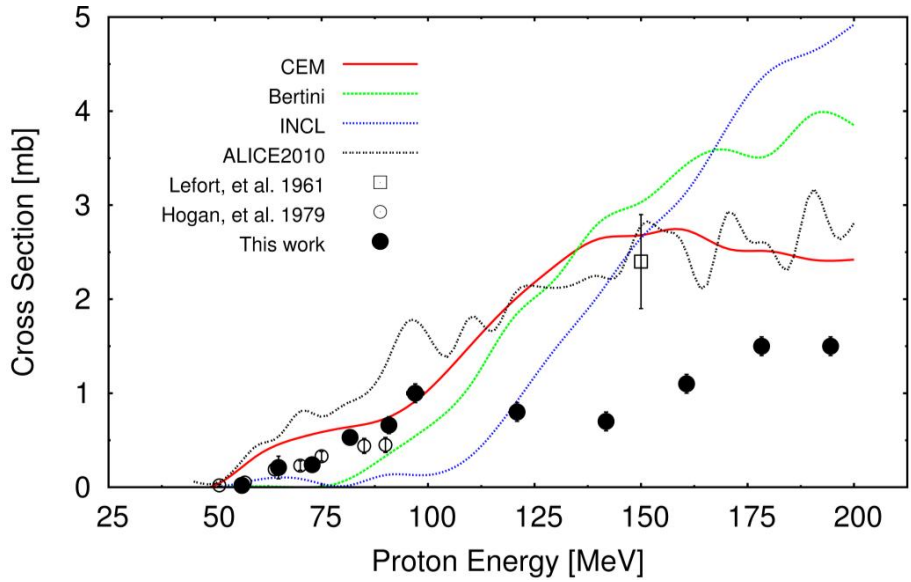

**FIG. 3.** Experimental and theoretical cumulative cross sections for the formation of $^{225}$Ra by the proton bombardment of thorium.

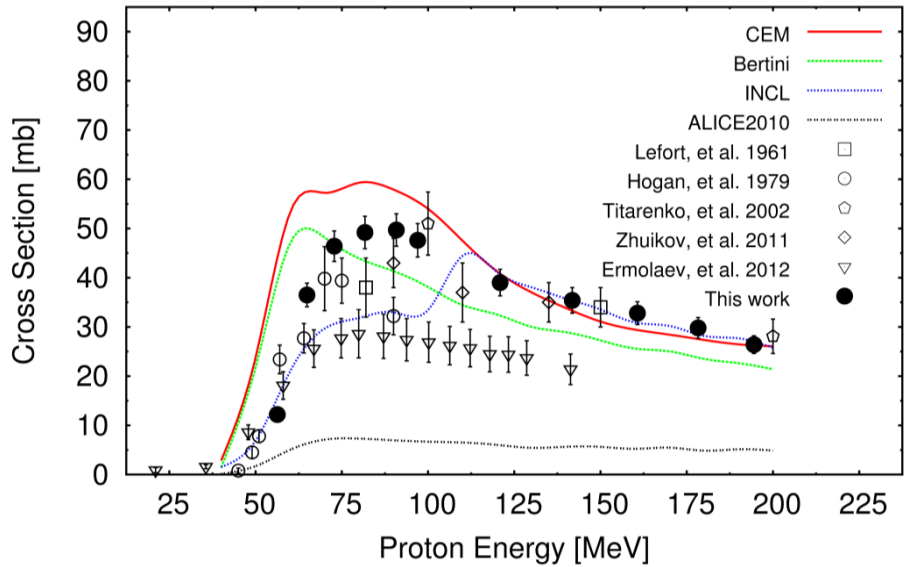

**FIG. 4.** Experimental and theoretical cumulative cross sections for the formation of $^{227}$Th by the proton bombardment of thorium.

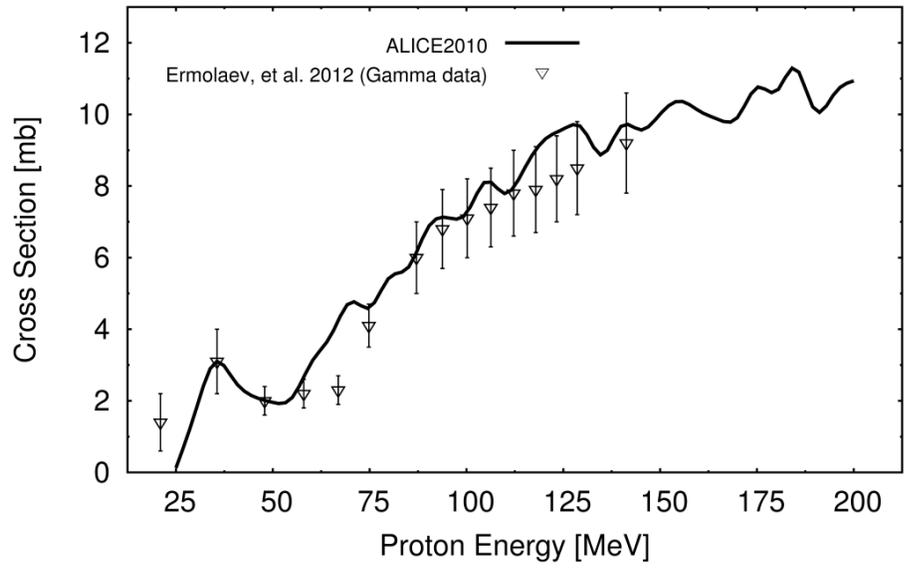

**FIG. 5.** Comparison of cumulative ALICE2010 cross section values and measurements obtained from gamma spectroscopy by Ermolaev et al., (2012) for the formation of $^{227}$Ac by the proton bombardment of thorium.

**Table 1**
Experimentally measured cumulative cross sections for the formation of $^{223,225}$Ra, $^{225}$Ac and $^{227}$Th.

| Proton Energy [MeV] | Cross Section [mb] | | | |
|---|---|---|---|---|
| | $^{225}$Ac | $^{223}$Ra | $^{225}$Ra | $^{227}$Th |
| 194.5 ± 0.3 | 17.5 ± 1.2 | 3.4 ± 0.3 | 1.5 ± 0.1 | 26.4 ± 1.8 |
| 178.3 ± 0.7 | 17.5 ± 1.2 | 3.0 ± 0.2 | 1.5 ± 0.1 | 29.8 ± 2.1 |
| 160.7 ± 1.0 | 17.0 ± 1.2 | 3.0 ± 0.3 | 1.1 ± 0.1 | 32.8 ± 2.3 |
| 141.8 ± 1.3 | 15.2 ± 1.1 | 2.3 ± 0.2 | 0.7 ± 0.1 | 35.4 ± 2.6 |
| 120.9 ± 1.6 | 11.4 ± 0.8 | 1.8 ± 0.2 | 0.8 ± 0.1 | 39.0 ± 2.7 |
| 97.0 ± 2.0 | 6.9 ± 0.5 | 2.1 ± 0.2 | 1.0 ± 0.1 | 47.6 ± 3.4 |
| 90.8 ± 0.4 | 6.0 ± 0.4 | 1.4 ± 0.3 | 0.66 ± 0.09 | 49.7 ± 3.3 |
| 81.7 ± 0.6 | 4.4 ± 0.3 | 0.81 ± 0.30 | 0.53 ± 0.07 | 49.2 ± 3.3 |
| 72.8 ± 0.7 | 4.1 ± 0.3 | 1.3 ± 0.3 | 0.24 ± 0.06 | 46.4 ± 3.1 |
| 64.9 ± 0.9 | 4.8 ± 0.3 | 0.54 ± 0.23 | 0.21 ± 0.12 | 36.5 ± 2.4 |
| 56.3 ± 1.1 | 4.1 ± 0.3 | 0.42 ± 0.12 | 0.02 ± 0.01 | 12.2 ± 0.8 |

**Table 2**
Production rates and projected yields from a 10-day irradiation of a 5 g/cm$^2$ natural thorium target at the IPF and BLIP. The beam current and energy range applicable at each facility are shown in parentheses.

| | IPF (250 µA, 93-72 MeV) | | BLIP (100 µA 195-183 MeV) | |
|---|---|---|---|---|
| | Production Rate[a] [µCi/µA·h] | Yield [Ci] | Production Rate[a] [µCi/µA·h] | Yield [Ci] |
| $^{225}$Ac | 33.1 | 1.4 | 115.6 | 2.0 |
| $^{223}$Ra | 6.8 | 0.3 | 18.8 | 0.3 |
| $^{225}$Ra | 2.6 | 0.1 | 6.7 | 0.1 |
| $^{227}$Th | 173.1 | 8.7 | 95.7 | 1.9 |
| $^{227}$Ac[b] | 0.04 | 0.003 | 0.09 | 0.002 |

[a] Instantaneous production rate, which does not account for decay
[b] Values calculated on the basis of the Ermolaev et al. data (see text)